\documentclass[letterpaper, 10 pt, conference]{ieeeconf}  % Comment this line out
\usepackage{amsmath}
\usepackage{amsfonts}
\usepackage{amssymb}
\usepackage{hyperref}
\IEEEoverridecommandlockouts                              % This command is only
\title{\LARGE \bf
A Quantum Optimal Control Problem with State Constrained Preserving Coherence}

\author{Nahid Binandeh Dehaghani and Fernando Lobo Pereira% <-this % stops a space

\thanks{SYSTEC - Research Center for Systems and Technologies, FEUP - Faculty of Engineering, Porto University, Rua Dr. Roberto Frias sn, i219, 4200-465 Porto, Portugal
        {\tt\small nahid@fe.up.pt},% 
        {\tt\small flp@fe.up.pt}}%
        }
\begin{document}
\maketitle
\thispagestyle{empty}
\pagestyle{empty}

%%%%%%%%%%%%%%%%%%%%%%%%%%%%%%%%%%%%%%%%%%%%%%%%%%%%%%%%%%%%%%%%%%%%%%%%%%%%%%%%
\begin{abstract}

In this work, we address the problem of maximizing fidelity in a quantum state transformation process controlled in such a way as to keep decoherence within given bounds. We consider a three-level $\Lambda$-type atom subjected to Markovian decoherence characterized by non-unital decoherence channels. We introduce fidelity as the performance index for the quantum state transformation process since the goal is to maximize the similarity of the final state density operator with the one of the desired target state. We formulate the quantum optimal control problem with state constraints where the later reflect the fact that the decoherence level remains within a pre-defined bound. Optimality conditions in the form of a Maximum Principle of Pontryagin in Gamkrelidze's form are given. These provide a complete set of relations enabling the computation of the optimal control strategy. This is a novel approach in the context of quantum systems.

\end{abstract}

%%%%%%%%%%%%%%%%%%%%%%%%%%%%%%%%%%%%%%%%%%%%%%%%%%%%%%%%%%%%%%%%%%%%%%%%%%%%%%%%
\section{INTRODUCTION}
The article is focused on the application of Maximum Principle of Pontryagin in the Gamkrelidze's form, \cite{arutyunov2011maximum} for state constrained optimal control problems arising in the control of quantum systems. To the best of our knowledge, this is the first instance of a formulation of  a optimal quantum control problem with state constraints, and, also, for which, this type of necessary conditions of optimality have been applied.
In this work, a three-level $\Lambda$-type atom subjected to Markovian decoherence characterized by non-unital decoherence channels is considered for which we preserve the quantum coherence within some bound specified by the state constraints. Then, by considering in the quantum state transformation process, as performance index, the quantum fidelity of density operators, we aim at maximizing the similarity of the desired target density operator with the final one.

With the increasing progress of quantum technology, the control of quantum phenomena has become a significant issue in a variety of areas such as quantum computing, quantum information processing, quantum optics, quantum chemistry, atomic and molecular physics, and nuclear magnetic resonance (NMR). Quantum control theory investigates how to drives quantum system dynamics by means of external controls so that tasks are performed, \cite{glaser2015training, d2021introduction, brif2010control, dong2010quantum},e.g., the application of quantum optimal control for quantum sensing, \cite{rembold2020introduction}, and quantum simulation, \cite{schaetz2013focus}.

Due to the extra noise and disturbances of closed-loop control methods in fragile quantum systems, the configuration of control law is considered without experimental feedback or measurement, as open-loop, in most quantum control protocols. Hence, optimal control theory (OCT) can be used as a strong tool to control quantum processes while minimizing a cost function according to the properties of quantum system, e.g., time optimal, \cite{khaneja2001time}, energy optimal, \cite{grivopoulos2004optimal}, etc. Moreover, in comparison with other control methods, OCT \cite{glaser2015training} has the key superiority of the very high versatility in the control problem formulation, which may include a wide variety of constraints, such as state, control, and mixed state-and-control and constraints. Necessary conditions of optimality for optimal control problem in the form of the Pontryagin’s Maximum Principle (PMP) has been already considered for a long time for quantum systems, \cite{peirce1988optimal,glaser2015training}. However, to the best of our knowledge, state constraints have not been considered. The reason for this relies on the fact that, in the presence of state constraints, the multipliers, in general, include Borel measures which make virtually impossible to use them in computationally efficient indirect methods based these optimality conditions. However, in \cite{arutyunov2011maximum,arutyunov2018remark}, a Maximum Principle of Pontryagin in the Gamkrelidze form was derived which under some regularity conditions, involve nontrivial multipliers which exhibit a degree of regularity that make indirect methods computationally effective.

The control of quantum decoherence, \cite{facchi2005control}, has attracted much attention as an important issue which has been remained unsolved yet. Loss of coherence means non-unitary evolutions, \cite{ai2006influences, niwa2002general}, leading to the loss of information and probability leakage towards the environment, which causes the lack of preservation of quantum superposition and entanglement, \cite{joos2013decoherence}. Under perfect isolation, a quantum system would indefinitely keep its coherence, however, the manipulation and investigation of an isolated quantum system is not possible. For instance, during the process of testing or manipulating a physical system in order to obtain numerical results, coherence is shared with the environment, and will be lost over periods of time. Until now, quantum presservation has been studied in several schemes such as quantum error-correction codes, \cite{shor1995scheme}, error-avoiding codes, \cite{duan1998prevention}, decoherence-free subspace, \cite{cappellaro2006principles}, dynamical decoupling, \cite{tasaki2004control}, tracking control, \cite{lidar2004stabilizing}. In addition, a feedback control strategy based on quantum weak measurements, \cite{zhang2010protecting}, has been investigated, however, the strategy was restricted to single-qubit systems and unital decoherence channels, while high-dimensional quantum systems have attracted much more attention in the applications of quantum control. Moreover, spontaneous emission, which can be modelled through non-unital decoherence channels, demonstrate properties different from unital decoherence channels and is typically used for decoherence modelling. Here, we take three-level quantum systems and spontaneous emission into account.

Amongst the main three quantum control problems, including quantum state preparation, transformation, and gate synthesis, the evolution of a quantum state from some given initial state into a desired target state is the basis of quantum computing, \cite{cory1997ensemble}, coherent control, \cite{warren1993coherent}, etc. The performance of such applications usually require a criterion, named fidelity, in order to be quantified. By means of comparison, fidelity provides a primary measure of similarity of two quantum states, for instance, in order to understand how accurately some given approximate calculation replicates some desired density operator, \cite{liang2019quantum}. However, fidelity can also be used for analysing quantum channels instead of pure states or density operators, \cite{gilchrist2005distance}. The concept of fidelity has been extensively used in quantum memories, \cite{chaneliere2018quantum}, quantum phase-space simulations, \cite{hillery1984distribution}, quantum teleportation, \cite{popescu1994bell}, etc.
In quantum teleportation, fidelity between the initial state and the final state is the standard measure used to quantify the effectiveness of the process, \cite{popescu1994bell}.

This work has been organized as follows: first we express the master equation indicating the mathematical model of a Markovian open quantum system and the system Hamiltonian. Then, we formulate the quantum coherence for non-unital decoherence channel and the ways to quantifying fidelity. We then present a novel formulation of the optimal control problem in the context of quantum physics with state constraints. In the next section, we present the solution to the proposed quantum control problem by means of Maximum Principle of Pontryagin in Gamkrelidze form for the first time in quantum context.

\section{System Dynamics}

According to Liouville equation we have
\begin{equation}\label{1}
\dot\rho (t)=-\frac{i}{\hbar }\left[ H\left( t \right),\rho ( t ) \right]
\end{equation}
where the quantum state is represented by the density operator $\rho$, and $\hbar$ is the reduced Planck’s constant. The system Hamiltonian $\displaystyle H(t)=H_{0}+H_C (u(t))=H_{0} +\sum_k u_k(t)H_k $ determines the controlled evolution, where $H_0$ and $H_k$ indicate the free and interaction Hamiltonians, respectively. The system control can be realized by a set of control functions $u_k(t)$ in $\mathbb{R}$, which is coupled to the quantum system via time independent interaction Hamiltonians $H_k$, $k=0,1,2,\dots$. The evolution of the Hamiltonian system is unitary
for closed quantum systems, so the spectrum of the quantum state is preserved. However, in many practical applications, the quantum system is open, and its evolution cannot be described by unitary transformations. The mathematical model of a Markovian open quantum system can be expressed by the master equation, \cite{breuer2002, popescu1994bellmarkovian, lindblad1976, alicki2007},
\begin{equation}\label{2}
	\dot\rho( t )=-\frac{i}{\hbar }\left[H( t),\rho ( t ) \right]+{\mathcal D}(\rho ( t ))
\end{equation}
where the dissipator $\mathcal{D}{{\rho }\left( t \right)}$ reflects the interactions between the system and the environment through different decoherence channels and is represented by
\begin{equation}\label{3}
\!\!\!\!\!{\mathcal D}(\rho( t ))\!=\!\sum_k \! \gamma_k \! \left( \! L_k\rho(t) L_k^\dagger \! - \! \frac{1}{2}L_k^\dagger L_k\rho (t)\!-\!\frac{1}{2}\rho (t)L_k^\dagger L_k \! \right)
\end{equation}
in which $L_k$ indicates the set of Lindblad operators, \cite{breuer2002}, and $\gamma_k$ is the damping rate.
By using the notation $vec\left( \left| \psi  \right\rangle \left\langle  \xi  \right| \right)=\left| \xi  \right\rangle \otimes \left| \psi  \right\rangle$, we can write \eqref{2} in vector space as
\begin{equation}\label{4}
    \frac{d}{dt}vec\left( \rho  \right)=\hat{\mathcal{L}}vec\left( \rho  \right)
\end{equation}
where
\begin{equation}\label{5}
\begin{aligned}
 &\! \! \!\hat{\mathcal{L}}=-i\left( I\otimes H-{{H}^{T}}\otimes I \right) \\ 
 &\!\!\!\! +\sum\limits_{k}{{{\gamma }_{k}}\left( L_{k}^{*}\otimes {{L}_{k}}-\frac{1}{2}I\otimes L_{k}^{\dagger }{{L}_{k}}-\frac{1}{2}{{\left( L_{k}^{\dagger }{{L}_{k}} \right)}^{T}}\otimes I \right)}\\ 
\end{aligned}
\end{equation}
in which $I$ indicates the identity matrix.

\section{Specification of the quantum control problem}
Consider a $\Lambda$-type atom subjected to Markovian decoherence. In ${\displaystyle \Lambda}$-type atoms, there are 3 states: ${\displaystyle |a\rangle }$, ${\displaystyle |b\rangle }$, and ${\displaystyle |e\rangle }$. In this type, ${\displaystyle |e\rangle }$ is at the highest energy level, while ${\displaystyle |a\rangle }$ and ${\displaystyle |b\rangle }$ are at lower levels. The ground states ${\displaystyle |a\rangle }$ and ${\displaystyle |b\rangle }$ can couple to the the excited state ${\displaystyle |e\rangle }$ with resonance frequencies (coupling constants) ${{\omega }_{1}}={\left( {{E}_{e}}-{{E}_{a}} \right)}/{\hbar }\;$ and ${{\omega }_{2}}={\left( {{E}_{e}}-{{E}_{b}} \right)}/{\hbar }$, respectively, in which ${{E}_{a}}$, ${{E}_{b}}$, and ${{E}_{e}}$ indicate the corresponding eigenvalues. ${{\omega }_{3}}$ can also be defined in a similar way to demonstrate the resonant frequency between ${\displaystyle |a\rangle }$ and ${\displaystyle |b\rangle }$ for the atom and laser field. The free Hamiltonian of such a system can be expressed as
\begin{equation}\label{6}
{{H}_{0}}=\frac{{{\omega }_{1}}}{3}\left( {{\rho }_{e}}-{{\rho }_{a}} \right)+\frac{{{\omega }_{2}}}{3}\left( {{\rho }_{e}}-{{\rho }_{b}} \right)+\frac{{{\omega }_{3}}}{3}\left( {{\rho }_{a}}-{{\rho }_{b}} \right)
\end{equation}
where ${{\rho }_{j}}=\left| j \right\rangle \left\langle  j \right|$, and the constant energy term ${\left( {{E}_{a}}+{{E}_{b}}+{{E}_{e}} \right)}/{3}$ is ignored. Since The excited state is not stable, two electrons in state ${\displaystyle |e\rangle }$ decay to the two ground states ${\displaystyle |a\rangle }$ and ${\displaystyle |b\rangle }$ at rates $\gamma_a$ and $\gamma_b$, and two kinds of emission are radiated, \cite{scully1999quantum}. By assuming that the decay process is Markovian, the decoherence channel can be characterized by the Lindblad operator ${{L}_{k}}=\left| k \right\rangle \left\langle  e \right|,$ $k=a,b$. in the dissipator equation \eqref{3}. In order to inhibit the decay process, we can apply a classical field such that drive the states between $|a\rangle$ and $|e\rangle$. By assuming that the transition dipole moments for the linearly polarized field are real, one can express the control field in the dipole approximation as $E\left( t \right)=u\left( t \right)\cos \left( {{\omega }_{d}}t+{{\phi }_{d}} \right)$, where $\omega_d$ and $\phi_d$ indicate the frequency and initial phase of the driving field. Hence, the control Hamiltonian is expressed as 
\begin{equation}\label{7}
    {{H}_{C}}\left( t \right)=u\left( t \right)\left( {{e}^{i{\phi }}}\left| a \right\rangle \left\langle  e \right|+{{e}^{-i{{\phi }}}}\left| e \right\rangle \left\langle  a \right| \right)\cos \left( {{\omega }}t \right)
\end{equation}

Here, decoherence channels are considered as non-unital, meaning that
\begin{equation}\label{8}
\mathcal{C}\left( \rho  \right)=\sqrt{\sum\limits_{i=1}^{2}{{{\left( tr\left( {{M}_{i}}\rho  \right) \right)}^{2}}}}
\end{equation}
where ${M_1}=\left| e \right\rangle \left\langle  a \right|+\left| a \right\rangle \left\langle  e \right|$ and ${M_2}=-i\left( \left| e \right\rangle \left\langle  a \right|-\left| a \right\rangle \left\langle  e \right| \right)$.
In order to preserve the coherence during the whole evolution, we impose the following constraint:
\begin{equation}\label{9}
 \alpha {{c}_{0}}\le \mathcal{C}\left( \rho  \right)\le {{c}_{0}}, \quad 0<\alpha <1
\end{equation}
In practice, any experimental quantum state  preparation or transformation is exposed to environmental noise and imperfections. Hence, one needs to quantify the degree of closeness for a pair of quantum states, i.e. the degree of accuracy of an approximate calculation for replication of some intended target density operator.
Given two density operators $\rho$ and $\sigma$, the fidelity is expressed as, \cite{liang2019quantum},
\begin{equation}\label{10}
{ \mathcal{F}(\rho ,\sigma )=\left(\operatorname {tr} {\sqrt {{\sqrt {\rho }}\sigma {\sqrt {\rho }}}}\right)^{2}}
\end{equation}
An equivalent expression for the fidelity may be written using the trace norm $\mathcal{F}(\rho ,\sigma )={\Big (}\operatorname {tr} |{\sqrt {\rho }}{\sqrt {\sigma }}|{\Big )}^{2}$ where the absolute value of an operator is here defined as ${\displaystyle |A|\equiv {\sqrt {A^{\dagger }A}}}$.

\section{Optimal Control Formulation}

Consider the following optimal control problem $(P_1)$

\begin{equation*}
\begin{split}
    \text{Minimize} \quad &-\mathcal{F}(\rho ,\sigma ) \\
    \text{subject to} \quad  &\dot{\rho}(t)=F\left( \rho(t),u(t) \right) \\
     & \rho\left( 0 \right)={{\rho}_{0}}\in {{\mathbb{R}}^{n}}\times{{\mathbb{R}}^{n}} \\
 & u\left( t \right)\in \mathcal{U}:=\left\{ u\in {{L}_{\infty }}:u\left( t \right)\in \Omega \subset {{\mathbb{R}}^{m}} \right\} \\
 & h\left( \rho\left( t \right) \right)\le 0 \quad \forall t\in \left[ 0,1 \right] \\
\end{split}
\end{equation*}
where $\dot{\rho }:=\displaystyle\frac{\partial \rho }{dt}$, $t\in \left[ 0,1 \right]$ determines the time variable, $\rho$ is the state variable in ${{\mathbb{R}}^{n}}\times{{\mathbb{R}}^{n}}$ supposed to satisfy the differential constraints $\dot{\rho}=F\left( \rho,u \right)$, and $\rho_0$ is the so-called initial quantum state. $u(\cdot)$ is the measurable bounded function termed as control. The inequality $h\left( \rho\left( t \right) \right)\le 0$ $\forall t\in \left[ 0,1 \right]$ defines the state constrains for the density operator. Here, we aim to maximize fidelity so that the density operator $\rho$ has the maximum overlap with the target $\sigma$. The fidelity criteria signifies a security level for quantum state transformation or the effectiveness of quantum gate synthesis.

Let $\left( {{\rho }^{*}},{{u}^{*}} \right)$ be the solution to the problem $(P_1)$. The pair $\left( {{\rho }^{*}},{{u}^{*}} \right)$ is termed as optimal if the value of fidelity is the maximum possible over the set of all feasible solutions. Then, there exist Lagrangian multipliers: $p$, $\mu$, and $\lambda$, where $p$ is an absolutely continuous function $p:\left[ 0,1 \right]\to {{\mathbb{R}}^{n}\times{\mathbb{R}}^{n}}$, $\mu$ is bounded variation, non-increasing  $\mu:\left[ 0,1 \right]\to {{\mathbb{R}}^{k}}$, such that $\mu(t)=cte$ on the time interval for which the state constraint is non-active, i.e., $\left\{ t\in \left[ 0,1 \right],  h\left( \rho\left( t \right) \right)<0 \right\}$, and $\lambda \ge 0$ is non-negative.
We also consider the following assumptions:
\begin{itemize}
    \item $h\left( \rho\left( t \right) \right)$ is $C^2$.
    \item $\left( {{\rho }^{*}},{{u}^{*}} \right)$ is a regular control process.
    \item The multiplier $\mu$ is continuous.
\end{itemize}
Since Gamkrelidze set of necessary conditions is used here, \cite{arutyunov2011maximum,arutyunov2017investigation}, the second-order derivative for $h\left( \rho\left( t \right) \right)$, which indicates state constraints, is required, \cite{arutyunov2018remark}.

\section{Maximum Principle}
For the indicated set of conditions, the extended Hamilton-Pontryagin function can be written as
\begin{equation}\label{11}
\mathcal{H}\left( \rho ,u,\pi ,\mu  \right)=tr\left( {{\left( \pi -\mu {{\nabla }_{\rho }}h \right)}^{\dagger }}F\left( \rho ,u \right) \right)
\end{equation}
in which $tr$ indicates the trace of matrix. According to Pontryagin's maximum principle, the optimal state trajectory $\rho^{*}$, optimal control $ u^{*}$, and corresponding Lagrange multiplier matrix $\pi^{*}$ must maximize the Hamiltonian $\mathcal{H}$ so that
\begin{equation}\label{12}
\mathcal{H}\left( {\rho }^{*}(t), u,{\pi}(t),\mu (t) \right)\le \mathcal{H}\left( {\rho }^{*}(t),{u}^{*}(t),{\pi}(t),\mu (t) \right)
\end{equation}
for all time $t\in [0,T]$ and for all permissible control inputs $u \in \mathcal{U}$. 

Additionally, the adjoint equation, and its terminal conditions imply that, Lebesgue a.e.,
\begin{equation}\label{13}
 ( -\dot\pi^\dagger(t),\dot\rho^*(t))=\nabla_{( \rho ,\pi)}  {\cal H}( \rho^*(t) ,u^*(t), \pi(t), \mu(t))
 \end{equation}
 \begin{equation}\label{14}
 \pi^\dagger( 1)=\nabla_\rho\mathcal{F}( \rho^*(1) ,\sigma ( 1 ))
\end{equation}
To satisfy the state constraint indicated in \eqref{9}, we defined
$h\left( \rho\left( t \right) \right)$ as the following
\begin{equation}\label{15}
h\left( \rho  \right)=\left( \begin{matrix}
   {{h}_{1}}\left( \rho  \right)  \\
   {{h}_{_{2}}}\left( \rho  \right)  \\
\end{matrix} \right)=\left( \begin{matrix}
   \mathcal{C}\left( \rho  \right)-{{c}_{0}}  \\
   \alpha {{c}_{0}}-\mathcal{C}\left( \rho  \right)  \\
\end{matrix} \right)
\end{equation}
For the sake of simplicity, we define $\bar{\mathcal{C}}\left( \rho  \right)={{\mathcal{C}}^{2}}\left( \rho  \right)$, and rewrite the constraints as
\begin{equation}\label{16}
\bar{h}\left( \rho  \right)=\left( \begin{matrix}
   {\bar{h}_{1}}\left( \rho  \right)  \\
   {\bar{h}_{{2}}}\left( \rho  \right)  \\
\end{matrix} \right)=\left( \begin{matrix}
   \bar{\mathcal{C}}\left( \rho  \right)-c_{0}^{2}  \\
   {{\alpha }^{2}}c_{0}^{2}-\bar{\mathcal{C}}\left( \rho  \right)  \\
\end{matrix} \right)\le 0
\end{equation}
Hence,
\begin{equation}\label{17}
{{\nabla }_{\rho }}\bar{h}\left( \rho  \right)=\left( \begin{matrix}
   {{\nabla }_{\rho }}\bar{\mathcal{C}}\left( \rho  \right)  \\
   -{{\nabla }_{\rho }}\bar{\mathcal{C}}\left( \rho  \right)  \\
\end{matrix} \right)
\end{equation}
in which ${{\nabla }_{\rho }}\bar{C}\left( \rho  \right)=2\sum\limits_{i=1}^{2}{\left( tr\left( {{M}_{i}}\rho  \right) \right){{M}_{i}}}$.

Noting that $\mu =\left( \begin{matrix}
   {{\mu }_{1}} & {{\mu }_{2}}  \\
\end{matrix} \right)
$, we can rewrite \eqref{11} as
\begin{equation}\label{18}
\!\!\!\!\mathcal{H}\!=\!tr\left(\! {{\left(\! \pi \! - \!\bar{\mu }\sum\limits_{i=1}^{2}{ tr\left( {{M}_{i}}\rho \! \right ) \!{{M}_{i}}}\! \right)}^{\dagger }}\!\!\left(\! -i\left[ H,\rho  \right]\!+\!\mathcal{D}\left( \rho  \right) \!\right)\! \right)
\end{equation}
where $\bar{\mu}=2(\mu_1-\mu_2)$.

From \eqref{13} we have
\begin{equation}\label{19}
    -{{\dot{\pi }}^{\dagger }}\!=\!\mathcal{G}\left( {{\pi }^{\dagger }} \right)\!-\!\bar{\mu} \sum\limits_{i=1}^{2}{\!\!{{M}_{i}}tr\!\left( \mathcal{G}\left( M_{i}^{\dagger } \right)\rho  \right)\!+\!tr\left( {{M}_{i}}\rho  \right)\!\mathcal{G}\left( M_{i}^{\dagger }\! \right)\! }
\end{equation}
where
\begin{equation}\label{20}
\mathcal{G}\left( \mathcal{X} \right)=-i\left[\mathcal{X},H\left( t \right) \right]+\mathcal{D}\left( \mathcal{X} \right)+{{\mathcal{D}}_{0}}\left( \mathcal{X} \right)
\end{equation}
in which
\begin{equation}\label{21}
{{\mathcal{D}}_{0}}\left( \mathcal{X} \right)=\sum\limits_{k}{{{\gamma }_{k}}\left( L_{k}^{\dagger }\mathcal{X}{{L}_{k}}-{{L}_{k}}\mathcal{X}L_{k}^{\dagger } \right)}
\end{equation}
Hence, by vectorization we have
\begin{equation}\label{22}
   - \frac{d}{dt}vec\left( {{\pi }^{\dagger }} \right)={\hat{\mathcal{L}}}'vec\left( {{\pi }^{\dagger }} \right)+vec\left(cte\left(\rho(t), \bar{\mu}(t) \right)\right)
\end{equation}
where
\begin{equation}\label{23}
    {\hat{\mathcal{L}}}'=\hat{\mathcal{L}}+\sum\limits_{k}{{\gamma }_{k}}\left(\left( L_{k}^{T}\otimes L_{k}^{\dagger } \right)+\left( L_{k}^{*}\otimes {{L}_{k}} \right)\right)
\end{equation}

The boundary condition \eqref{14} at the final time is obtained by computing, \cite{Dehaghani2022},
\begin{equation}\label{24}
\pi^\dagger( 1)=\nabla_\rho\left( tr\sqrt{\sqrt{\rho(1) }\sigma(1) \sqrt{\rho (1) }}\right)^2
\end{equation}
for which by means of Taylor expansion at the point $\rho = I$, we have
\begin{equation}\label{25}
\sqrt{\rho }=\sum_{k=0}^\infty\frac{1}{k!}\frac{d^k}{d\rho^k}\sqrt{\rho}_{|\rho = I}( \rho -I)^k
\end{equation}
which, in turn, under some conditions, the application of the Cayley–Hamilton theorem yields, for a certain coefficients $\alpha_k$, $k=0,\ldots, n-1$, being $n$ the dimension of the square matrix $\rho$,
$$ \sqrt{\rho }=\sum_{k=0}^{n-1}\alpha_k( \rho -I)^k.$$ Thus, by using matricial calculus, we obtain
\begin{equation}\label{26}
\begin{aligned}
&\pi^{\dagger}( 1)=2 tr\sqrt{\rho(1)\sigma(1)}\nabla_\rho( tr\sqrt{\rho(1) \sigma(1) })\\
 &= 2 tr\sqrt{\rho(1) \sigma(1) }\sum\limits_{k=0}^{n-1}{\alpha }_{k}\sum_{i=0}^{k-1}\bar \rho(1)^i\sqrt{\sigma ( 1 )}\bar \rho(1)^{k-i-1}
\end{aligned}
\end{equation}
where $ \bar \rho(1)= \rho(1)-I$.

\section{Application of the Pontryagin Maximum Principle}

Let denote the excited state as
\begin{equation}\label{27}
\rho_e=\left( \begin{matrix}
   1 & 0 & 0  \\
   0 & 0 & 0  \\
   0 & 0 & 0  \\
\end{matrix} \right)    
\end{equation}
and the ground states as 
\begin{equation}\label{28}
\rho_a=\left( \begin{matrix}
   0 & 0 & 0  \\
   0 & 1 & 0  \\
   0 & 0 & 0  \\
\end{matrix} \right), \quad \rho_b=\left( \begin{matrix}
   0 & 0 & 0  \\
   0 & 0 & 0  \\
   0 & 0 & 1  \\
\end{matrix} \right)     
\end{equation}
The drift Hamiltonian for such system is expressed as
\begin{equation}\label{29}
\begin{aligned}
 {{H}_{0}}&=\sum\limits_{j=1}^{3}{{{E}_{j}}{{\rho }_{j}}} =0.8{{\rho }_{e}}+0.5{{\rho }_{a}}+0.4{{\rho }_{b}} \\ 
 &=\left( \begin{matrix}
   0.8 & 0 & 0  \\
   0 & 0.5 & 0  \\
   0 & 0 & 0.4  \\
\end{matrix} \right)\\
\end{aligned}
\end{equation}
and the control Hamiltonian is
\begin{equation}\label{30}
    {{H}_{C}}\left( t \right)=u\left( t \right)\left( \begin{matrix}
   0 & {{e}^{-i{\phi }}} & 0  \\
   {{e}^{i{{\phi }}}} & 0 & 0  \\
   0 & 0 & 0  \\
\end{matrix} \right)\cos \left( {{\omega }}t \right)
\end{equation}
From \eqref{3}, we have 
\begin{equation}\label{31}
\begin{aligned}
  & \mathcal{D}\left( \rho \left( t \right) \right)\!=\!10^{-1}\!\left(\! {{L}_{1}}\rho \left( t \right)L_{1}^{\dagger }-\frac{1}{2}L_{1}^{\dagger }{{L}_{1}}\rho \left( t \right)-\frac{1}{2}\rho \left( t \right)L_{1}^{\dagger }{{L}_{1}} \!\right) \\ 
 & +10^{-3}\left( {{L}_{2}}\rho \left( t \right)L_{2}^{\dagger }-\frac{1}{2}L_{2}^{\dagger }{{L}_{2}}\rho \left( t \right)-\frac{1}{2}\rho \left( t \right)L_{2}^{\dagger }{{L}_{2}} \right) \\ 
\end{aligned}
\end{equation}
in which
\begin{equation*}
 {{L}_{1}}=\left( \begin{matrix}
   0 & 0 & 0  \\
   1 & 0 & 0  \\
   0 & 0 & 0  \\
\end{matrix} \right)  , \quad {{L}_{2}}=\left( \begin{matrix}
   0 & 0 & 0  \\
   0 & 0 & 0  \\
   1 & 0 & 0  \\
\end{matrix} \right)\cdot
\end{equation*}
In this example, the initial state reflects the fact that the initial population distribution is mostly on the levels $\rho_a$ and $\rho_e$. Hence, as it can be seen from \eqref{31}, the relaxation rate $\gamma_2$ is considerably smaller in comparison with $\gamma_1$, so the three-level system can be considered as a two-level one. The coherence of the system is expressed by
\begin{equation}\label{32}
    \bar{\mathcal{C}}\left( \rho  \right)={{\left( tr\left( {{M}_{1}}\rho  \right) \right)}^{2}}+{{\left( tr\left( {{M}_{2}}\rho  \right) \right)}^{2}}
\end{equation}
where ${{M}_{1}}=\left( \begin{matrix}
   0 & 1 & 0  \\
   1 & 0 & 0  \\
   0 & 0 & 0  \\
\end{matrix} \right)$ and ${{M}_{2}}=\left( \begin{matrix}
   0 & -i & 0  \\
   i & 0 & 0  \\
   0 & 0 & 0  \\
\end{matrix} \right)$. Suppose that $\rho_0=\rho_e$ is the initial state, meaning that the system remains on the excited state. Since we consider the effectiveness of our method for any pure state, we present the desired target state by
\begin{equation}\label{33}
    \sigma \left( 1 \right)=\left( \begin{matrix}
   0 & 0 & 0  \\
   0 & {{\sin }^{2}}\beta  & \sin \beta \cos \beta   \\
   0 & \sin \beta \cos \beta  & {{\cos }^{2}}\beta   \\
\end{matrix} \right)\
\end{equation}
for $\beta \in [0,2\pi)$.

Now, we present a time discretized computational scheme to solve the optimal control problem associated with $(P_1)$ by using an indirect method based on the Maximum Principle.
Let $N$ be the number of discrete time subintervals. Given the smooth properties of the problems data, let us consider a uniform discretization. Thus, we consider the $N$ points $\displaystyle t_m = \frac{m}{N}$ for $ m = 0,\ldots, N-1$ and denote the value of any function $f(t_m)$ by $f_m$. We consider both the system dynamics, and the adjoint differential equation approximated by a first order Euler approximation. Higher order methods yield better approximations but our option is made to keep the presentation simple. Let $j=0,\ldots$ be the iterations counter, and we denote the $j^{th}$ iteration of the function $f$ at time $t_m$ by $f_m^j$. 
\\
From now on the control is considered as $V_{m}^{j}=\left( u_{m}^{j},{{\phi }^{j}},{{\omega }^{j}} \right)$, in which $u$ depends on time, however, $\phi$ and $\omega $ modify in each iteration.
\vspace{0.2cm}

The proposed algorithm is as follows:
\begin{itemize}
\item[] \hspace{-.5cm}{\bf Step 1 - Initialization}. \vspace{.3cm}

\noindent Let $j=0$, for $m=1,\ldots,N-1$ in \eqref{30},
\begin{itemize}
\item[] Initialize $V_{m}^{j}=\left( u_{m}^{j},{{\phi }^{j}},{{\omega }^{j}} \right)$
\end{itemize}
\vspace{.3cm}
\item[] \hspace{-.5cm}{\bf Step 2 - Computation of the state trajectory}. \vspace{.1cm}
\vspace{.3cm}
\\
For $m=0,\dots,N$ 
\begin{itemize}
\item[] Vectorize $\rho_0^j$.
\item[] Compute
$$vec\left( \rho _{m}^{j} \right)={{e}^{\sum\limits_{k=0}^{m-1}{\frac{1}{N}\hat{\mathcal{L}}\left( V_{k}^{j} \right)}}}vec\left( \rho _{0}^{j} \right),$$
\item[] and obtain
$$\rho _{m}^{j}=ve{{c}^{-1}}\left( \rho _{m}^{j} \right)\cdot$$
\end{itemize}
\vspace{.3cm}

\item[] \hspace{-.5cm}{\bf Step 3 - Computation of the adjoint trajectory}.
\vspace{.3cm}
\\
Compute $\pi_N^j$ by using \eqref{26} with $\rho _N^j$ computed in Step 2.
\newline
Initialize $\mu^j_{1,N}=0$ and $\mu^j_{2,N}=0$.
\newline
Obtain $vec(\pi_N^j)$ and $vec(cte(\rho_m^j,\bar{\mu}_m^j))$.
\vspace{.1cm}

For $m=N-1$ to $0$, 
\begin{itemize}
\item[] Compute
\begin{equation*}
\begin{aligned}
vec\left( \pi _{m}^{j} \right)&={{e}^{\sum\limits_{k=m}^{N-1}{\frac{1}{N}{\hat{\mathcal{L}}}'\left( V_{k}^{j} \right)}}}vec\left( \pi _{N}^{j} \right) \\ 
&\!\!\!\!\!\!\!\!\!\! +\sum\limits_{k=m}^{N-1}{\frac{1}{N}{{e}^{\sum\limits_{l=k}^{N-1}{\frac{1}{N}{\hat{\mathcal{L}}}'\left( V_{l}^{j} \right)}}}vec\left( cte\left( \rho _{m}^{j}, \bar{\mu}_m^j \right) \right)} \\ 
\end{aligned}
\end{equation*}
\item[] Obtain 
$$\pi _{m}^{j}=ve{{c}^{-1}}\left( vec\left( \pi _{m}^{j} \right) \right)$$
\item[] Obtain $$\bar{\mu }_{m-1}^{j}=2\left( \mu _{1,m-1}^{j}-\mu _{2,m-1}^{j} \right)$$ by checking
\begin{itemize}
    \item[*] If $\mathcal{\bar{C}}(\rho_m^j)<c_0^2$, then 
    $$\mu^j_{1,m-1}=\mu^j_{1,m} $$ 
    and if $\mathcal{\bar{C}}(\rho_m^j)>\alpha^2 c_0^2$, then 
    $$\mu^j_{2,m-1}=\mu^j_{2,m} $$
\end{itemize}  
or
\begin{itemize}   
    \item[*] If $\mathcal{\bar{C}}(\rho_m^j)=c_0^2$, then $$\mu^j_{1,m-1}=\mu^j_{1,m}+\frac{1}{N}{{\nabla }_{\rho }}\bar{\mathcal{C}}\left( \rho _{m}^{j} \right)$$ and if $\mathcal{\bar{C}}(\rho_m^j)=\alpha^2 c_0^2$, then $$\mu^j_{2,m-1}=\mu^j_{2,m}-\frac{1}{N}{{\nabla }_{\rho }}\bar{C}\left( \rho _{m}^{j} \right)$$
\end{itemize}

\end{itemize}

\vspace{0.3cm}
\item[] \hspace{-.5cm}{\bf Step 4 - Computation of the Pontryagin Hamilton function}\vspace{.3cm}
\\
For $m=0,\dots,N-1$, let $\mathcal{H}_m^j(u,\omega,\phi)$ as
$$
\!\!\!\!\!tr\left(\! {{\left(\! \pi^j_m \! - \!{\bar{\mu}^j_m }\sum\limits_{i=1}^{2}{ tr\left( {{M}_{i}}\rho \right){{M}_{i}}}\! \right)}^{\dagger }}\!\!\left(\! -i\left[ H^j_m,\rho^j_m  \right]\!+\!\mathcal{D}\left( \rho  \right) \!\right)\! \right)
$$
where $\displaystyle H(u,\omega,\phi,t_m)$, from \eqref{29} and \eqref{30}, is given by
\vspace{0.3cm}
$$
   \left( \begin{matrix}
   0.8 & u{{e}^{-i{\phi }}}cos(\omega t_m) & 0  \\
   u{{e}^{i{{\phi }}}}cos(\omega t_m) & 0.5 & 0  \\
   0 & 0 & 0.4  \\
\end{matrix} \right)
$$

\vspace{2Cm} 
\item[] \hspace{-.5cm}{\bf Step 5: Update the control function.}\vspace{.1cm}
\vspace{.3cm} 
\\
For $m=0,\ldots, N-1$, compute the values $u_m^{j+1}$, $\omega^{j+1}$, and $\phi^{j+1}$ that maximize the map
$$ (u,\omega,\phi) \to \mathcal{H}_m^j(u,\omega,\phi).$$

\vspace{.2cm} 
\item[] \hspace{-.5cm}{\bf Step 6: Stopping test}.\vspace{.1cm}

\noindent For given small positive numbers $\varepsilon_\omega$, $\varepsilon_\phi $, and $\varepsilon_u$ (tolerance errors), check whether all the following inequalities hold:
\begin{eqnarray*}
&&\hspace{1.5cm} | \omega^j-\omega^{j+1}|<\varepsilon_\omega \\
&& \hspace{1.5cm}| \phi^j-\phi^{j+1}|<\varepsilon_\phi \\
&& \hspace{.5cm}  \max_{m=0,\ldots,N-1}\{|u_m^j-u_m^{j+1} |\}<\varepsilon_u 
\end{eqnarray*}

\smallskip

\noindent If yes, let
$$\omega^*= \omega^{j+1}$$
$$\phi^*= \phi^{j+1}$$
and for $m=0,\ldots,N-1$,
$$u^*(t_m) = u_m^{j+1}$$
and exit the algorithm. 

\smallskip 

\noindent Otherwise, let $ j=j+1$, go to {\bf Step 2}.
\end{itemize}

\vspace{0.3cm}

\section{Conclusion}
In this article, we have shown the application of Maximum Principle of Pontryagin in Gamkrelidze's form in maximizing fidelity in a quantum state transformation process controlled in such a way as to preserve coherence within given bounds while, simultaneously, maximizing the fidelity in the transfer of the state variable. The context of the matrix-valued probability density function, between two given state values with the dynamics of the system expressed by the Lindblad master equation was adapted. We introduced a complete set of equations enabling the computation of optimal control strategy in the presence of state constraints, which is a novel approach in the context of quantum systems. As an example, we took into account a three-level $\Lambda$-type atom subjected to Markovian decoherence characterized by non-unital decoherence channels. We finally introduced an algorithm to solve the proposed problem.

%%%%%%%%%%%%%%%%%%%%%%%%%%%%%%%%%%%%%%%%%%%%%%%%%%%%%%%%%%%%%%%%%%%%%%%%%%%%%%%%
\section{ACKNOWLEDGMENTS}
The authors acknowledge the support of FCT for the grant 2021.07608.BD, the ARISE AL, Ref. LA/P/0112/2020, SYSTEC-Base, Ref. UIDB/00147/2020, and Programmatic, Ref. UIDP/00147/2020, and also the support of projects SNAP, Ref. NORTE-01-0145-FEDER-000085, and MAGIC, Ref. PTDC/EEI-AUT/32485/2017, COMPETE2020-POCI and FCT/MCTES (PIDDAC).

%%%%%%%%%%%%%%%%%%%%%%%%%%%%%%%%%%%%%%%%%%%%%%%%%%%%%%%%%%%%%%%%%%%%%%%%%%%%%%%%

\end{document}